\documentclass{ws-ijseke}
\usepackage{cite}
\usepackage{svg}
\usepackage{amsmath,amssymb,amsfonts}
\usepackage{algorithmic}
\usepackage{graphicx}
\usepackage{textcomp}
\usepackage{xcolor}
\usepackage{booktabs} 
\usepackage{graphicx}
\usepackage{xcolor}
\usepackage{xspace}
\usepackage[normalem]{ulem}
\useunder{\uline}{\ul}{}
\usepackage{lscape}
\usepackage{rotating}
\usepackage{supertabular}
\usepackage{longtable}
\usepackage{enumerate}
\usepackage{svg}
\usepackage{amsmath}
\usepackage{colortbl}
\usepackage{subcaption}
\usepackage{hhline}
\usepackage{makecell}
\usepackage{caption}
\usepackage{hyperref}
\usepackage{enumitem}
\usepackage{tcolorbox}
\usepackage{tabularx}
\usepackage{lipsum}
\usepackage{listings}
\usepackage{tabularx,ragged2e}
\usepackage{verbatim}
\usepackage{fancyvrb}
\usepackage{balance}
\usepackage{multirow}
\begin{document}

\markboth{B. Zhang et al.}
{Demystifying Practices, Challenges and Expected Features of Using GitHub Copilot}

%
\catchline{}{}{}{}{}
%

\title{Demystifying Practices, Challenges and Expected Features of Using GitHub Copilot}

\author{Beiqi Zhang, Peng Liang\footnote{Corresponding author.}, Xiyu Zhou}

\address{School of Computer Science, Hubei Luojia Laboratory, Wuhan University\\
Wuhan 430072, China\\\
\email{\{zhangbeiqi, liangp, xiyuzhou\}@whu.edu.cn}
}

\author{Aakash Ahmad}

\address{School of Computing and Communications, Lancaster University Leipzig\\
Leipzig 04109, Germany\\
a.ahmad13@lancaster.ac.uk
}

\author{Muhammad Waseem}

\address{School of Computer Science, Hubei Luojia Laboratory, Wuhan University\\
Wuhan 430072, China\\\
m.waseem@whu.edu.cn
}

\maketitle

\begin{history}
\received{Day Month Year)}
\revised{(Day Month Year)}
\accepted{(Day Month Year)}
\end{history}

\begin{abstract}
With the advances in machine learning, there is a growing interest in AI-enabled tools for autocompleting source code. GitHub Copilot, also referred to as the ``AI Pair Programmer'', has been trained on billions of lines of open source GitHub code, and is one of such tools that has been increasingly used since its launch in June 2021. However, little effort has been devoted to understanding the practices, challenges, and expected features of using Copilot in programming for auto-completed source code from the point of view of practitioners. To this end, we conducted an empirical study by collecting and analyzing the data from Stack Overflow (SO) and GitHub Discussions. More specifically, we searched and manually collected 303 SO posts and 927 GitHub discussions related to the usage of Copilot. We identified the programming languages, Integrated Development Environments (IDEs), technologies used with Copilot, functions implemented, benefits, limitations, and challenges when using Copilot. The results show that when practitioners use Copilot: (1) The major programming languages used with Copilot are \textit{JavaScript} and \textit{Python}, (2) the main IDE used with Copilot is \textit{Visual Studio Code}, (3) the most common used technology with Copilot is \textit{Node.js}, (4) the leading function implemented by Copilot is \textit{data processing}, (5) the main purpose of users using Copilot is \textit{to help generate code}, (6) the significant benefit of using Copilot is \textit{useful code generation}, (7) the main limitation encountered by practitioners when using Copilot is \textit{difficulty of integration}, and (8) the most common expected feature is that Copilot \textit{can be integrated with more IDEs}. Our results suggest that using Copilot is like a double-edged sword, which requires developers to carefully consider various aspects when deciding whether or not to use it. Our study provides empirically grounded foundations that could inform software developers and practitioners, as well as provide a basis for future investigations on the role of Copilot as an AI pair programmer in software development.
\end{abstract}


\section{Introduction}
\label{sec:introduction}
Large Language Models (LLMs) and Machine Learning (ML) for autocompleting source code are becoming more and more popular in software development. LLMs nowadays incorporate powerful capabilities for Natural Language Processing (NLP) \cite{austin2021program}, and ML approaches have been widely applied to source code using a variety of new tools to support software development \cite{allamanis2018survey}, which makes it possible to use LLMs to synthesize code in general-purpose languages \cite{austin2021program}. Recently, NLP-based code generation tools have come into the limelight, with generative pre-trained language models trained on large corpus of code in an attempt to provide reasonable auto-completion of the source code when programmers write code \cite{hammond2021empirical}. Released in June 2021, GitHub Copilot has recently emerged as an ``AI pair programmer'', which is powered by OpenAI Codex and suggests code or entire functions in IDEs as a plug-in \cite{githubcopilot} to help developers achieve code auto-completion in development.

Although the emergence of AI-assisted programming tools has empowered practitioners in their software development efforts, there is little evidence and lack of empirically-rooted studies (e.g, \cite{hammond2021empirical}, \cite{bird2022taking}, \cite{madi2023association}) on the role of AI-assisted programming tools in software development. The existing studies such as \cite{nguyen2022empirical} and \cite{yetistiren2022assessing} primarily focus on the correctness and understanding of the code suggested by Copilot, and little is known about the practices, challenges, and expected features of using Copilot during programming and software development activities for the developers and users of Copilot. \textbf{To ameliorate this gap}, we conducted this study that collects data from Stack Overflow (SO) and GitHub Discussions to get practitioners' perspectives on using Copilot during software development. 
While Bird \textit{et al.} investigated Copilot users' initial experiences of how they would use Copilot, as well as what challenges they encountered by three studies \cite{bird2022taking}, our study explored the practices, challenges, and expected features of using Copilot by analyzing the data from developer communities. The emergence of Copilot has shifted the paradigm of pair programming, and it is challenging for software development teams to adopt this approach and tool on a large scale \cite{bird2022taking}. Our work intends to provide developers and users of GitHub Copilot with comprehensive insights on the practices, purposes, and expected features of using Copilot.

\textbf{The contributions of this work}: (1) we identified the programming languages, IDEs, and technologies used with Copilot; (2) we provided the functions implemented by Copilot, the purposes, benefits, limitations/challenges, and expected features of using Copilot; and (3) we collected and added more data related to Copilot from SO and GitHub Discussions (134 posts and 272 discussions, leading to 303 posts and 927 discussions in total) for purposes of enhancing the external validity of our study results.

This paper is an extension of our previous conference paper \cite{zhang2023practices} published in the Proceedings of the 35th International Conference on Software Engineering and Knowledge Engineering (SEKE 2023), and the following additions show how we extended the previous work: (1) we extended our dataset by including the latest data from Stack Overflow and GitHub Discussions formulated before June 18th, 2023 (303 SO posts and 927 GitHub discussions in total); we explored and reported the purposes of using GitHub Copilot in RQ1.4; (3) we investigated and reported the expected features of users about GitHub Copilot from practitioners' perspectives in RQ2.3; and (4) we provide more implications based on the study results.

The structure of the paper: Section \ref{sec:relatedWork} surveys the related work, and Section \ref{sec:researchDesign} presents the research design of this study. Section \ref{sec:results} provides the study results, which are further discussed in Section \ref{sec:implications}. The potential threats to validity are clarified in Section \ref{sec:threats} and Section \ref{sec:conclusions} concludes this work with future directions.

\section{Related Work}
\label{sec:relatedWork}
\subsection{Analyzing the Code Generated Using Copilot}
Several studies focused on the security issues of Copilot. Sandoval \textit{et al.} \cite{gustavo2022security} conducted a user study to investigate the impact of programming with LLMs that support Copilot. Their results show that LLMs have a positive impact on the correctness of functions, and they did not find any decisive impact on the correctness of safety. Several studies focused on the quality of the code generated by Copilot. Imai \cite{imai2022github} compared the effectiveness of programming with Copilot versus human programming, and found that the generated code by Copilot is inferior than human-written code. Yetistiren \textit{et al.} \cite{yetistiren2022assessing} assessed the quality of generated code by Copilot in terms of validity, correctness, and efficiency. Their empirical analysis shows Copilot is a promising tool. Madi \textit{et al.} \cite{madi2023association} focused on readability and visual inspection of Copilot generated code. Through a human experiment, their study highlights that programmers should beware of the code generated by tools. Wang \textit{et al.} \cite{wang2023practitioners} collected practitioners' expectations on code generation tools through a mixed-methods approach. They found that effectiveness and code quality is more important than other expectations. Several studies focused on the limitations and challenges in Copilot assisted programming.

\subsection{Capabilities and Limitations of Copilot}
Several studies have explored the capabilities and limitations of GitHub Copilot. Dakhel \textit{et al.} \cite{dakhel2022asset} explored Copilot's capabilities through empirical evaluations, and their results suggest that Copilot shows limitations as an assistant for developers. Nguyen and Nadi \cite{nguyen2022evalution} conducted an empirical study to evaluate the correctness and comprehensibility of the code suggested by Copilot. Their findings revealed that Copilot’s suggestions for different programming languages do not differ significantly, and they identified potential shortcomings of Copilot, like generating complex code. Bird \textit{et al.} \cite{bird2022taking} conducted three studies to understand how developers use Copilot and their findings indicated that developers spent more time assessing suggestions by Copilot than doing the task by themselves. Sarkar \textit{et al.} \cite{sarkar2022artificial} compared programming with Copilot to previous conceptualizations of programmer assistance to examine their similarities and differences, and discussed the issues that might arise in applying LLMs to programming.

Compared to the existing work (e.g., \cite{yetistiren2022assessing}, \cite{dakhel2022asset}), our work intends to understand the practices, challenges, and expected features of using Copilot by exploring various aspects of Copilot's usage.

\section{Research Design}
\label{sec:researchDesign}
\begin{figure}[htbp]
	\centering
	\includegraphics[width=0.8\linewidth]{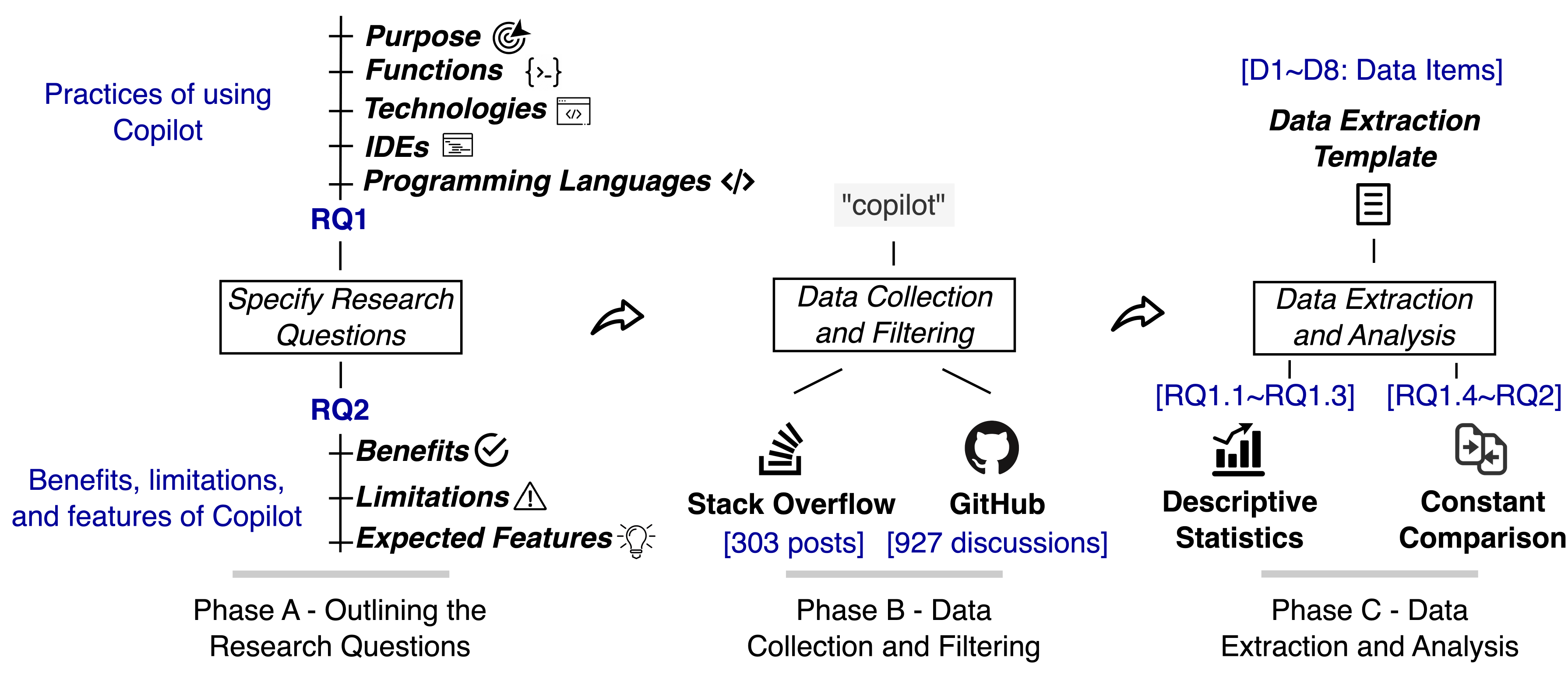}
	\caption{Overview of the research process}
	\label{fig:Overview of research process}
\end{figure}

The goal of this study is to understand the practices, benefits and challenges, and expected features of using GitHub Copilot from the point of view of practitioners in the context of Copilot related SO posts and GitHub discussions. We conducted and reported this exploratory study by following the guidelines for empirical studies in software engine ring proposed in \cite{runeson2009guidelines}. We formulated two Research Questions (RQs) with eight sub-RQs that contribute to the goal of this study, as presented in Section \ref{subsec:researchQuestions} with their rationale. The overview of the research process is shown in Figure \ref{fig:Overview of research process} and detailed in Section~\ref{subsec:datacollectionandfitering} and Section~\ref{subsec:dataextractionandanalysis}.

\subsection{Phase A - Outlining the Research Questions}
\label{subsec:researchQuestions}
We aimed to explore the characteristics of practices, challenges, and expected features of using GitHub Copilot by answering the RQs in the following:

\noindent \textbf{RQ1: Programming Languages, IDEs, Technologies, Functions, and Purposes of Using GitHub Copilot}\\
\noindent \textbf{RQ1.1: What programming languages are used with GitHub Copilot?}\\
\noindent \emph{Rationale:} In software development, the role of programming languages is fundamental - enabling programmers to translate requirements by writing source code using a specific programming language such as Java or Python. This RQ aims to collect the programming languages that developers tend to use with Copilot.\\
\noindent \textbf{RQ1.2: What IDEs are used with GitHub Copilot?}\\
\noindent \emph{Rationale:} Copilot is a third-party plug-in used in IDEs. This RQ aims to identify the IDEs frequently used with Copilot. The answers of this RQ can help developers choose which IDE to use when they code with Copilot.\\
\noindent \textbf{RQ1.3: What technologies are used with GitHub Copilot?}\\
\noindent \emph{Rationale:} When writing code, programmers need to employ certain technologies to complete the development. This RQ aims to investigate the technologies that can be used with Copilot (e.g., frameworks), and the answers of this RQ can help developers to choose the technologies when they use Copilot.\\
\noindent \textbf{RQ1.4: What functions are implemented by using GitHub Copilot?}\\
\noindent \emph{Rationale:} Copilot can complete entire functions according to users' comments. This RQ aims to provide a categorization of the functions that can be implemented by Copilot, and the answers of this RQ can provide developers guidance when implementing functions by using Copilot.\\
\noindent \textbf{RQ1.5: What are the purposes of using GitHub Copilot?}\\
\noindent \emph{Rationale:} Users may use Copilot for different purposes. This RQ aims to explore what are users using Copilot for, and the answers of this RQ can help us get a better understanding of how people are using Copilot.\\

\noindent \textbf{RQ2: Benefits, Limitations \& Challenges, and Expected Features of Using GitHub Copilot}\\
\noindent \textbf{RQ2.1: What are the benefits of using GitHub Copilot?}\\
\noindent \emph{Rationale:} Using Copilot to assist programming can bring many benefits (e.g., reducing the workload of developers). This RQ aims to collect the advantages brought to the development by applying Copilot.\\
\noindent \textbf{RQ2.2: What are the limitations \& challenges of using GitHub Copilot?}\\
\noindent \emph{Rationale:} Although using Copilot to assist in writing code can help developers with their programming activities, there are still restrictions and problems when using Copilot. This RQ aims to collect and identify the limitations and challenges practitioners may experience when using Copilot. The answers of this RQ can help practitioners make an informed decision when deciding whether to code with the help of Copilot.\\
\noindent \textbf{RQ2.3: What are the expected features of users about GitHub Copilot?}\\
\noindent \emph{Rationale:} Since its released on June, 2021, GitHub Copilot has become increasingly popular among programmers and has been developed into a mature automated code-completion tool. However, there are still some features Copilot does not provide but users want to have. This RQ aims to investigate the expected features that users expect about GitHub Copilot. The answers of this RQ can give suggestions to the development team of GitHub Copilot, guiding them to make Copilot an AI-enabled coding tool that can better meet users' needs.\\

\subsection{Phase B - Data Collection and Filtering}
\label{subsec:datacollectionandfitering}
This study focuses on understanding the practices, challenges, and expected features of using Copilot collected from SO and GitHub Discussions. SO is a popular software development community and has been widely used by developers to ask and answer questions as a Q\&A platform. GitHub Discussions is a feature of GitHub used to support the communication among the members of a project. Different from SO, GitHub Discussions can provide various communication intentions, not just question-answering (e.g., a discussion can report errors or discuss the potential development of a software project) \cite{hata2022github}, from which the data can be complementary to the data from SO. Besides, GitHub Discussions can provide a center of a community knowledge base connected with other artifacts in a project \cite{hata2022github}. Therefore, we decided to use SO and GitHub Discussions as the data sources to answer our RQs, and we conducted the searches for both SO and GitHub Discussions on June 18th, 2023. We conducted the data filtering manually because the numbers of posts from SO and discussions from GitHub Discussions are not large, which could be completed in an acceptable effort and time by manual work. Another reason for using a manual approach is that conducting data filtering using an automatic approach would lead to false positive filtering results and thus make the filtering results inaccurate, which will negatively affect the findings we obtain.

\textbf{For SO}, ``\textit{copilot}'' is used as the term to search the posts related to Copilot. After searching, we got a total of 714 posts that include the search term ``\textit{copilot}''. The term ``\textit{copilot}'' may appears more than once in a post, so there may be duplicates in the URL collection of these retrieved posts. After removing the duplicates, we ended up with 678 posts with unique URLs. We conducted the data To manually label posts related to Copilot, we conducted a pilot data labelling by two authors with 10 retrieved SO posts. Specifically, the inclusion criterion is that the post must provide information referring to Copilot. We calculated the Cohen's Kappa coefficient \cite{jacob1960coefficient} to measure the consistency of labelled posts, which is 0.773, thus indicating a decent agreement between the two coders. After excluding the irrelevant posts in the search results, we finally got 303 Copilot related SO posts.

\textbf{For GitHub Discussions}, GitHub discussions are organized according to categories. After exploring the categories on GitHub Discussions, we found the ``Copilot'' category which contains the feedback, questions, and conversations about Copilot \cite{githubdisucssions} under the ``GitHub product categories''. Since all the discussions under the ``Copilot'' category are related to Copilot, we then included all the discussions under the ``Copilot'' category as related discussions to extract data. The number of discussions related to Copilot is 927.

\subsection{Phase C - Data Extraction and Analysis}
\label{subsec:dataextractionandanalysis}
To answer the RQs in Section \ref{subsec:researchQuestions}, similar to Data Collection and Filtering in Section~\ref{subsec:datacollectionandfitering}, we manually extracted the data items listed in Table \ref{Data items extracted and their corresponding RQs}. The first and third authors conducted a pilot data extraction independently with 10 SO posts and 10 GitHub discussions randomly selected from the 303 SO posts and 927 GitHub discussions. The second author was involved to discuss with the two authors and came to an agreement if any disagreements were found during the pilot. After the pilot, the criteria for data extraction were determined: (1) for all the data items listed in Table \ref{Data items extracted and their corresponding RQs}, they will be extracted and counted only if they are explicitly mentioned by developers that they were used with Copilot; (2) if the same developer repeatedly mentioned the same data item in an SO post or a GitHub discussion, we only counted it once. In a post or discussion, multiple developers may mention Copilot related data items, resulting in situation that the total number of instances of certain data item extracted may be greater than the number of posts and discussions. The first and third authors further extracted the data items from the filtered posts and discussions according to the extraction criteria, marked uncertain parts, and discussed with the second author to reach a consensus. Finally, the first author rechecked all the extraction results by the two authors from the filtered posts and discussions to ensure the correctness of the extracted data.

\begin{table}[htbp]
\scriptsize
\caption{Data items extracted and their corresponding RQs}
\label{Data items extracted and their corresponding RQs}
\begin{tabular}{p{0.6cm}<{\centering}p{3.3cm}p{6.2cm}p{0.8cm}}
\hline
\textbf{\#} & \textbf{Data Item}                     & \textbf{Description}                                                & \textbf{RQ}   \\ \hline
D1          & Programming language                   & \textit{Programming languages used with Copilot}                    & RQ1.1    \\ \hline
D2          & IDE                                    & \textit{IDEs used with Copilot}                                     & RQ1.2    \\ \hline
D3          & Technology                             & \textit{Technologies used with Copilot}                             & RQ1.3    \\ \hline
D4          & Function                               & \textit{Functions implemented by Copilot}                           & RQ1.4    \\ \hline
D5          & Purpose                                & \textit{Intentions of using Copilot}                                & RQ1.5    \\ \hline
D6          & Benefit                                & \textit{Benefits brought by using Copilot}                          & RQ2.1    \\ \hline
D7          & Limitation and challenge               & \textit{Restrictions and difficulties when using Copilot}           & RQ2.2    \\ \hline
D8          & Expected feature                       & \textit{Features that users want Copilot to provide}                & RQ2.3    \\ \hline
\end{tabular}
\end{table}


\begin{table}[htbp]
\scriptsize
\caption{Data items and their analysis methods}
\label{Data Items and the data analysis methods used for the research questions}
\begin{tabular}{m{0.6cm}<{\centering}m{4.5cm}m{5cm}m{0.8cm}<{\centering}}
\hline
\textbf{\#} & \textbf{Data Item}                         & \textbf{Data Analysis Method}                                & \textbf{RQ}  \\ \hline
D1          & Programming language                       & Descriptive statistics~\cite{christopher2017interpreting}    & RQ1.1 \\ \hline
D2          & IDE                                        & Descriptive statistics                                       & RQ1.2 \\ \hline
D3          & Technology                                 & Descriptive statistics                                       & RQ1.3 \\ \hline
D4          & Function                                   & Constant comparison~\cite{stol2016grounded}                  & RQ1.4 \\ \hline
D5          & Purpose                                    & Constant comparison                                          & RQ1.5 \\ \hline
D6          & Benefit                                    & Constant comparison                                          & RQ2.1 \\ \hline
D7          & Limitation and challenge                   & Constant comparison                                          & RQ2.2 \\ \hline
D8          & Expected Feature                           & Constant comparison                                          & RQ2.3 \\ \hline
\end{tabular}
\end{table}

\subsubsection{Analyze Data}
For RQ1.1, RQ1.2, and RQ1.3, we used descriptive statistics \cite{christopher2017interpreting} to analyze and present the results. For RQ1.4, RQ1.5, and RQ2, we conducted a qualitative data analysis by applying the Constant Comparison method \cite{glaser1965the}. We constantly compared each part of the data (e.g., emergent codes) to explore differences and similarities in the extracted data to form categories \cite{core2006lillemor}. Note that for answering RQ1.4, we categorized the functions (D4) based on developers' discussions, i.e., developers' descriptions of the mentioned functions. Firstly, the first and the third authors coded the filtered posts and discussions with the corresponding data items listed (see Table \ref{Data items extracted and their corresponding RQs}). Secondly, the first author reviewed the coded data by the third author to make sure the extracted data were coded correctly. Finally, the first author combined all the codes into higher-level concepts and turned them into categories. After that, the second author examined the coding and categorization results, in which any divergence was discussed till the three authors reached an agreement. The data analysis methods with their corresponding data items and RQs are listed in Table \ref{Data Items and the data analysis methods used for the research questions}. The data analysis results are provided in \cite{replpack}.

\section{Results}
\label{sec:results}
\begin{figure*}[htbp]
	\centering
	\includegraphics[width=1.1\linewidth]{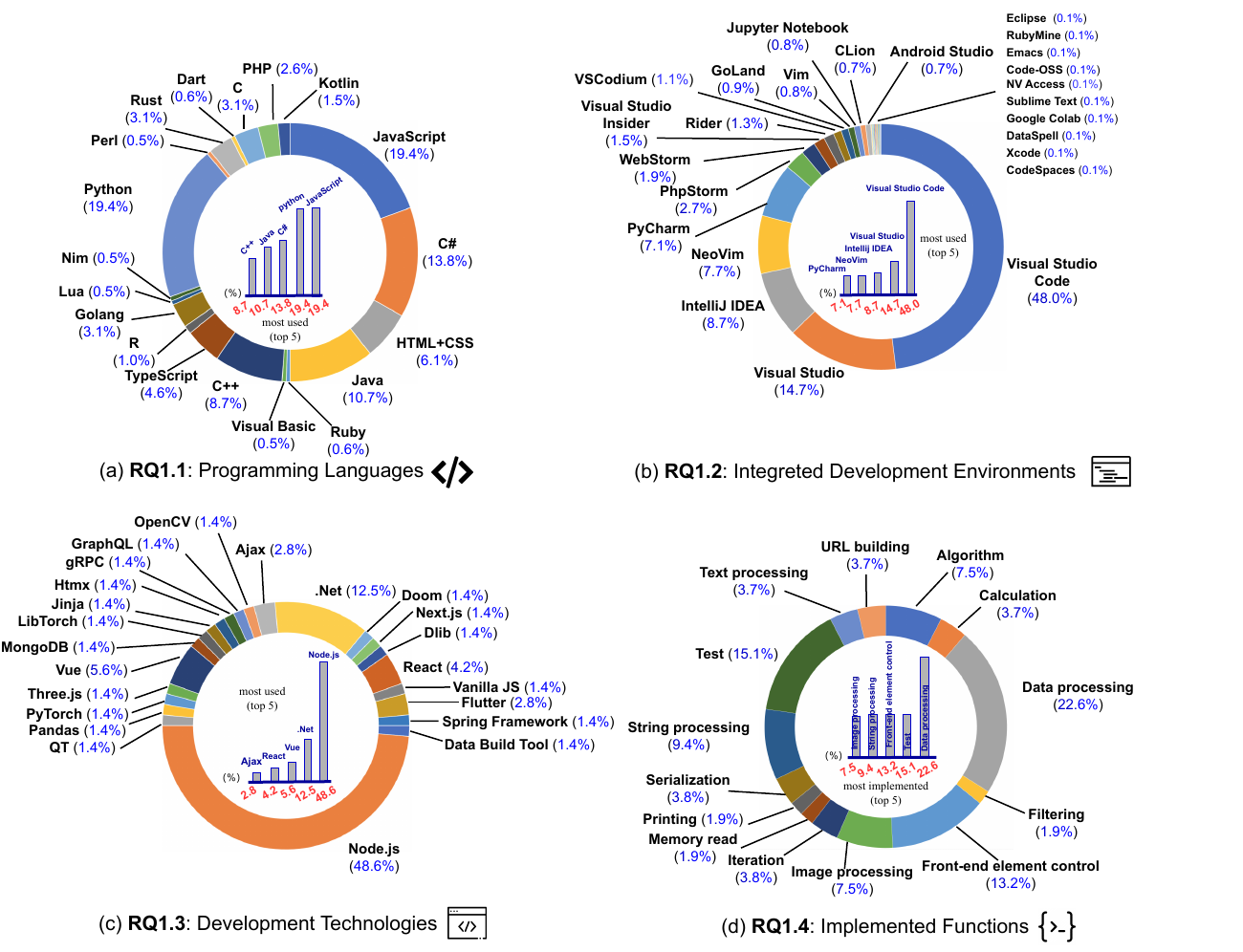}
	\caption{Programming languages, IDEs, technologies, and implemented functions of using Copilot (results of RQ1.1 to RQ1.4)}
	\label{fig:Results of RQ1.1-RQ1.4}
\end{figure*}

In this section, the study results of RQ1.1 to RQ1.4 are visualized in Fig \ref{fig:Results of RQ1.1-RQ1.4}, and the results of RQ1.5 and RQ2.1 to RQ2.3 are provided in Table \ref{Purposes of using Copilot}, \ref{Benefits of using Copilot}, \ref{Limitations and challenges of using Copilot}, and \ref{Expected features of users about Copilot}.

\subsection{RQ1: Programming Languages, IDEs, Technologies, and Functions of Using GitHub Copilot}
\noindent \textbf{RQ1.1: What programming languages are used with GitHub Copilot?}\\
Figure \ref{fig:Results of RQ1.1-RQ1.4}a lists the 19 programming languages used with Copilot, in which \textit{JavaScript} and \textit{Python} are the most frequently used ones, both accounting for one fifth. Besides, developers often write \textit{C\#} and \textit{Java} code when using Copilot, as one practitioner mentioned ``\textit{the GitHub Copilot extension is enabled in my VS 2022 C\# environment}'' (GitHub \#14115). \textit{HTML+CSS}, \textit{TypeScript}, \textit{Golang}, \textit{C}, \textit{Rust}, \textit{PHP}, and \textit{Kotlin} were used with Copilot between 3\textasciitilde12 times each (1.5\% to 6.1\%). The rest programming languages (e.g., \textit{Perl}, \textit{Ruby}), and \textit{Visual Basic} which are not popular, were mentioned only once with Copilot.


\noindent \textbf{RQ1.2: What IDEs are used with GitHub Copilot?}\\
Figure \ref{fig:Results of RQ1.1-RQ1.4}b shows 25 types of IDEs that are used with Copilot. \textit{Visual Studio Code} is the dominant IDE, accounting for 48.0\%. When first released, Copilot only worked with \textit{Visual Studio Code} editor, and it is expected that \textit{Visual Studio Code} is the IDE most often used with Copilot. Mainstream code editors, including \textit{Visual Studio}, \textit{IntelliJ IDEA}, \textit{NeoVim}, and \textit{PyCharm} are also occasionally used, account for 38.2\% in total. The remaining IDEs were rarely mentioned by developers, and one possible reason is that there are often integration issues when using Copilot within them according to the results of RQ2.2.

\noindent \textbf{RQ1.3: What technologies are used with GitHub Copilot?}\\
Figure \ref{fig:Results of RQ1.1-RQ1.4}c presents 23 technologies used with Copilot. We find that these identified technologies include frameworks, APIs, and libraries. \textit{Node.js}, whose proportion is more than 45\%, is one of the most popular back-end runtime environments for \textit{JavaScript}, which is also the most frequently used language with Copilot (see the results of RQ1.1), thus it is reasonable that \textit{Node.js} is the major technology used with Copilot. In addition, \textit{.NET} which works for Web development, and \textit{Vue}, \textit{React}, \textit{Flutter}, and \textit{Ajax} which are frameworks for front-end development, were mentioned less often compared to \textit{Node.js}. The rest of the identified technologies, many of which relate to machine learning (e.g., \textit{Pandas}, \textit{Dlib}, and \textit{OpenCV}) or front-end development (e.g., \textit{Htmx}, \textit{Vanilla JS}, and \textit{Next.js)}, are rarely used with Copilot, and each of them appears only once.

\noindent \textbf{RQ1.4: What functions are implemented by using GitHub Copilot?}\\
Figure \ref{fig:Results of RQ1.1-RQ1.4}d shows 14 functions implemented by using Copilot. The main function implemented by Copilot is \textit{data processing}, indicating that developers tend to make use of Copilot to write functions working with data. Besides, \textit{test} (15.1\%) and \textit{front-end element control} (13.2\%) are the functions that account for more than 10\% besides \textit{data processing}. When implementing functions, developers also use Copilot to code \textit{string processing}, \textit{image processing}, and \textit{algorithm}, among which \textit{image processing} and \textit{algorithm} account for the same, i.e., 7.5\%. The rest types of functions are seldom implemented by using Copilot, which are mentioned by developers twice or once.

\noindent \textbf{RQ1.5: What are the purposes of users using GitHub Copilot?}\\
Table \ref{Purposes of using Copilot} shows nine purposes of using Copilot. 43.3\% of the developers from SO and GitHub Discussions indicated that they used Copilot \textit{to help generate code} they needed. Another 9 developers (15.0\%) said that they wanted \textit{to try out the functionality of Copilot}, so they downloaded the tool and used it. \textit{To fix bugs} is the third most popular reason for developers to use Copilot. When developers found their code did not work and they could not fix the bugs by themselves, they would turn to Copilot for help, as one developer mentioned in the post ``\textit{I've been working with copilot and chatgpt for days trying to get the code to work by fixing it}'' (SO \#76498229). \textit{To improve coding ability} and \textit{to provide ideas for writing code} have the same percentage, 6.7\%. Some developers used Copilot to help them learn knowledge related to coding \textit{to improve coding ability}, for examples, one developer said that ``\textit{I am trying to learn js}'' (GitHub \#6947) when he applied for free use of Copilot. Some developers did not want to directly use the code suggested by Copilot, and they only wanted to refer to the suggestions of Copilot to provide them with ideas on how to solve their problems. Copilot can be used \textit{for educational purposes} and \textit{for research purposes} as well. Users also use Copilot to help them \textit{generate code comments}, as one developer mentioned that he used ``\textit{Copilot in Visual Studio 2022 to document code}'' (SO \#76070342), but the percentage of this purpose is only 5.0\%. Besides, two developers (3.3\%) made use of Copilot \textit{to check the code} because they could not find why the code did not work.

\begin{table*}[htbp]
\caption{Purposes of using GitHub Copilot (results of RQ1.5)}
\scriptsize
\label{Purposes of using Copilot}
\begin{tabular}{p{3.2cm}p{6.6cm}p{0.7cm}<{\centering}p{0.6cm}<{\centering}}
\hline
\textbf{Purpose}                                         & \textbf{Example}        & \textbf{Count}   & \textbf{\%} \\ \hline
To help generate code                                    & \textit{GitHub CoPilot was able to fill in the code I needed after I wrote the steps in comments} (SO \#71905508)                                                                                                           & 26               & 43.3\%      \\ \hline
To try out the functionality of Copilot                  & \textit{I use VIM and IntelliJ on a daily basis, and I recently installed VS Code to try out the newest "Copilot Chat" features} (SO \#76349751)                                                                        & 9                & 15.0\%       \\ \hline
To fix bugs                                              & \textit{spent a month trying to get it to work on my own and after a week of trying to get chatgpt or copilot to fix the bugs I'm all out of ideas} (SO \#76266095)                                                      & 7                & 11.7\%       \\ \hline
To improve coding ability                                & \textit{I'm improving my rusty skills lately and saw (in some Copilot suggestions) the question mark operator used as a prefix of variables} (SO \#74008676)                                                         & 4                & 6.7\%        \\ \hline
To provide ideas for writing code                        & \textit{With the input from jps (AES is actually OK for encrypted tokens, but not signed) and Github Copilot I came up with a working solution using HMAC-SHA256} (SO \#72812667)                                        & 4                & 6.7\%        \\ \hline
For educational purposes                                 & \textit{For the past year I have used and taught my students how they could benefit from co-pilot while coding} (GitHub \#19410)                                                                                            & 3                & 5.0\%        \\ \hline
To generate code comments                                & \textit{Using ChatGPT and Copilot, I've commented it to understand its functionality.} (SO \#75624961)                                                                                                                      & 3                & 5.0\%        \\ \hline
To check the code                                        & \textit{I've checked myself by stepping through, I've checked with GitHub Copilot, I've checked with ChatGPT, and they all say this is correct.} (SO \#76311798)                                                          & 2                & 3.3\%        \\ \hline
For research purposes                                    & \textit{I am working on a scientific study testing how Copilot will effect the academic setting} (GitHub \#8324)                                                                                                            & 2                & 3.3\%        \\ \hline
\end{tabular}
\end{table*}

\begin{table*}[htbp]
\caption{Benefits of using GitHub Copilot (results of RQ2.1)}
\scriptsize
\label{Benefits of using Copilot}
\begin{tabular}{p{3.4cm}p{6.3cm}p{0.8cm}<{\centering}p{0.6cm}<{\centering}}
\hline
\textbf{Benefit}                                         & \textbf{Example}        & \textbf{Count}   & \textbf{\%} \\ \hline
Useful code generation                                   & \textit{I find myself writing a lot of tests, and Copilot is excellent at helping with writing repetitive tests} (GitHub \#9282)                                                                                         & 28               & 45.9\%      \\ \hline
Faster development                                       & \textit{I really enjoy using it , it reduce programming time} (GitHub \#17382)                                            
                                                                                   & 10                & 16.4\%       \\ \hline
Better code quality                                      & \textit{it's faster and simpler to your solution} (SO \#68418725)                                                                
                                                                                   & 7                & 11.5\%       \\ \hline
Good adaptation to users' code patterns                  & \textit{GitHub copilot adapt to your coding practices} (SO \#69740880)                                                          
                                                                                   & 4                & 6.6\%        \\ \hline
Better user experience                                   & \textit{Since copilot works totally different compared to all the other products out there, it is a lot more fun to use and does not annoy me like some other AI systems} (GitHub \#7254)                               & 4                & 6.6\%        \\ \hline
Free for students                                        & \textit{If you are a student you can sign up for the GitHub Student Pack, which gives a lot of benefits, one being copilot for free} (GitHub \#31494)                                                                   & 3                & 4.9\%        \\ \hline
Powerful code interpretation and conversion functions    & \textit{Does Copilot have the code explanation feature or something similar? It does! some active members were given beta access.} (GitHub \#38089)                                                                   & 2                & 3.3\%        \\ \hline
Frequent updates to provide more features                & \textit{Keep in mind that there are updates to the plugin very frequently, so there's still hope} (SO \#70428218)                                                                                                        & 1                & 1.6\%        \\ \hline
Strong integration capability                            & \textit{GitHub is supporting more editors} (GitHub \#6858)                                                                                                                                                               & 1                & 1.6\%        \\ \hline
Ease of study and use                                    & \textit{when using this plugin, can study at a relatively low cost} (GitHub \#8028)                                                                                                                                      & 1                & 1.6\%        \\ \hline
\end{tabular}
\end{table*}
\subsection{RQ2: Benefits, Limitations \& Challenges, and Expected Features of Using GitHub Copilot}
\noindent \textbf{RQ2.1: What are the benefits of using GitHub Copilot?}\\
Table \ref{Benefits of using Copilot} highlights 10 benefits of using Copilot. Most developers mentioned that they used Copilot for \textit{useful code generation}, which reduced their workload and gave them help when they have no idea about how to write code. Programming with Copilot also brings \textit{faster development}, as one discussion remarked, Copilot ``\textit{saves developers a lot of time}'' (GitHub \#35850). Meanwhile, \textit{better code quality} can be obtained by using Copilot. Compared to the code written by developers themselves, the code suggested by Copilot is usually shorter and more correct, as one developer said, ``\textit{often Copilot is smarter than me}'' (SO \#74512186). Copilot can use machine learning models to learn code style of developers, so as to offer \textit{good adaptation to users' code patterns}. Four developers mentioned that Copilot can give them \textit{better user experience} than other AI-assisted programming tools, for example, one developer stated that ``\textit{Copilot works totally different compared to all the other products out there, it is a lot more fun to use and does not annoy me like some other AI systems}'' (GitHub \#7254), without providing the names of the other products.

\noindent \textbf{RQ2.2: What are the limitations \& challenges of using GitHub Copilot?}\\
Table \ref{Limitations and challenges of using Copilot} lists 15 limitations and challenges of using Copilot. Most developers pointed out the \textit{difficulty of integration} between Copilot and IDEs or other plug-ins. After Copilot was installed in developers' IDEs, certain plug-ins did not work and Copilot may conflict with some shortcut settings of the editors. Moreover, Copilot cannot be successfully integrated with some IDEs as Copilot does not support these editors yet. Due to the instability of Copilot servers, no support for proxies, and access restriction of some regions, developers may have \textit{difficulties of accessing Copilot}. The code suggested by Copilot has constraints as well, and sometimes it just offers few solutions, which are not enough for users, which brings \textit{limitation to code generation}, as one developer said ``\textit{multiple solution is too little}'' (GitHub \#37304). Practitioners also complained about the \textit{poor quality of generated code} by Copilot. Some practitioners said that ``\textit{GitHub Copilot suggest solutions that don't work}'' (SO \#73701039), and some practitioners found that when the code files became larger, the quality of the code suggested by Copilot ``\textit{becomes unacceptable}'' (GitHub \#9282). When using Copilot, developers pay much attention to \textit{code privacy threat} as well. They were worried that Copilot may use their code information without permission. Contrary to the developers who mentioned that Copilot gave them a \textit{better user experience} than other AI-assisted programming tools, some practitioners said they had an \textit{unfriendly user experience} when coding with Copilot. Compared to the results of our previous work \cite{zhang2023practices}, the number of \textit{difficulty of subscription} increases significantly. This may be caused by the restrictions for free users, as one user complained ``\textit{so looks like people with a free plan are stuck on a rate-limit for now with no way out}'' (GitHub \#43893).

\begin{scriptsize}
\begin{center}
\begin{longtable}{p{3.4cm}p{6.3cm}p{0.8cm}<{\centering}p{0.6cm}<{\centering}}
\caption{Limitations and challenges of using Copilot (results of RQ2.2)}\\
\scriptsize
\label{Limitations and challenges of using Copilot} \\ \hline
\textbf{Limitation \& Challenge}                   & \textbf{Example}        & \textbf{Count}   & \textbf{\%} \\ \hline 
Difficulty of integration                          & \textit{Copilot only works with VSCode, VSCodium is not supported at the moment} (GitHub \#14837)                                                                                                                        & 114               & 28.1\%      \\ \hline
Difficulty of accessing Copilot                    & \textit{I cannot connect to the GitHub account and the Copilot server in VSCode, also cannot use the Copilot plugin} (SO \#74398521)                                                                                     & 69               & 17.0\%      \\ \hline
Limitation to code generation                      & \textit{Copilot is limited to around 1000 characters in the response} (GitHub \#15122)                                           
                                                                             & 48               & 11.8\%       \\ \hline
Poor quality of generated code                     & \textit{Github Copilot suggest solutions that don't work} (SO \#73701039)                                                             
                                                                             & 36               & 8.9\%       \\ \hline
Code privacy threat                                & \textit{Copilot does collect personal data so just take precaution when working in private repos} (GitHub \#7163)                                                                                                        & 29               & 7.1\%       \\ \hline
Unfriendly user experience                         & \textit{I had the same problem today, an amazing tool with poor user experience} (GitHub \#8468)                                                                                                                         & 25               & 6.2\%       \\ \hline
Difficulty of subscription                         & \textit{My copilot subscription suddenly stopped. Tried log out and in. Never have reply on support ticket over 10 days} (GitHub \#36190)                                                                                & 22                & 5.4\%       \\ \hline
High pricing                                       & \textit{it is obvious that no one in South America will pay that price, it is too expensive} (GitHub \#24594)                                                                                                            & 14               & 3.4\%       \\ \hline
Lack of customization                              & \textit{My question is about setting up shortcuts in Visual Studio Code VSCode for GitHub Copilot Labs.} (SO \#73564811)                                                                                                 & 13                & 3.2\%       \\ \hline
Difficulty of understanding the generated code     & \textit{I really do not understand this enough, and have no idea half of what this code does honestly. It was written by Copilot.} (SO \#72282605)                                                                  & 12                & 3.0\%       \\ \hline
Hard to configure                                  & \textit{Keep getting "Your Copilot experience is not fully configured, complete your setup" in Visual Studio 2022} (GitHub \#19556)                                                                                      & 10                & 2.5\%       \\ \hline
No edition for organizations                       & \textit{Currently, Copilot is only available for individual user accounts and organizations aren't able to purchase/manage Copilot for their members just yet} (GitHub \#32775)                                     & 8                & 2.0\%       \\ \hline
Show loading                                       & \textit{I am not sure what is causing this but while editing files within Visual Studio, I am periodically locking up with the following dialog showing} (SO \#73682137)                                              & 3                & 0.7\%       \\ \hline
Challenge of not providing outdated suggestions    & \textit{making sure that the tool does not provide outdated suggestions would still be a challenge} (SO \#72554382)                                                                                                      & 2                & 0.5\%       \\ \hline
Need of basic programming knowledge                & \textit{It is useless if you do not understand the programming language or the task you want to do} (GitHub \#35850)                                                                                                     & 1                & 0.2\%       \\ \hline
\end{longtable}
\end{center}
\end{scriptsize}

\begin{table*}[htbp]
\caption{Expected features of users about Copilot (results of RQ2.3)}
\scriptsize
\label{Expected features of users about Copilot}
\begin{tabular}{p{10cm}p{0.8cm}<{\centering}p{0.6cm}<{\centering}}
\hline
\textbf{Expected Feature}                                    & \textbf{Count}   & \textbf{\%} \\ \hline
Can be integrated with more IDEs                             & 32               & 28.8\%      \\ \hline
Allow customization of shortcuts for suggestions             & 12               & 10.8\%      \\ \hline
Give suggestions when requested                              & 8                & 7.2\%      \\ \hline
A team version                                               & 7                & 6.3\%      \\ \hline
Support access proxies                                       & 7                & 6.3\%      \\ \hline
Allow customization of the format of generated code          & 5                & 4.5\%      \\ \hline
Accept the needed part of the suggestions                    & 4                & 3.6\%      \\ \hline
Compatible with other code generation tools                  & 4                & 3.6\%      \\ \hline
Allow setting filters for suggestions                        & 3                & 2.7\%      \\ \hline
Allow self-signed certificates                               & 3                & 2.7\%      \\ \hline
Can be used with more development frameworks                 & 2                & 1.8\%      \\ \hline
Ability to turn off data collection                          & 2                & 1.8\%      \\ \hline
Suggestions in IDE UI can be configured                      & 2                & 1.8\%      \\ \hline
Code explanation                                             & 2                & 1.8\%      \\ \hline
Free for certain type of users                               & 2                & 1.8\%      \\ \hline
Provide more suggestions at a time                           & 2                & 1.8\%      \\ \hline
Provide more complete suggestions                            & 2                & 1.8\%      \\ \hline
Ability to draw UML digrams                                  & 1                & 0.9\%      \\ \hline
Enable a dialog to accept or deny suggestions                & 1                & 0.9\%      \\ \hline
A version for CLI (Command-Line Interface)                   & 1                & 0.9\%      \\ \hline
A on-premises version                                        & 1                & 0.9\%      \\ \hline
Ability to select the training sources for suggestions       & 1                & 0.9\%      \\ \hline
Can be used in remote servers                                & 1                & 0.9\%      \\ \hline
Provide a getting started guide                              & 1                & 0.9\%      \\ \hline
Provide a security rating for generated code                 & 1                & 0.9\%      \\ \hline
Remind users when it has no suggestions                      & 1                & 0.9\%      \\ \hline
Show acceptance rate of suggestions                          & 1                & 0.9\%      \\ \hline
View the code-related data shared by Copilot                 & 1                & 0.9\%      \\ \hline
Disable notification sounds of suggestions                   & 1                & 0.9\%      \\ \hline
\end{tabular}
\end{table*}

\noindent \textbf{RQ2.3: What are the expected features of users about GitHub Copilot?}\\
Table \ref{Expected features of users about Copilot} presents 29 features that users expected to use about Copilot. The most often mentioned feature by users is that they hope Copilot \textit{can be integrated with more IDEs} (28.8\%). According to the results of RQ2.2, the dominant limitation and challenge of using Copilot is \textit{difficulty of integration} as Copilot cannot be used in some editors. It is then reasonable that many users want Copilot to support more IDEs. 10.8\% users remarked that they wanted Copilot to \textit{allow customization of shortcuts for suggestions}, as one post wrote ``\textit{Github CoPilot should give us an option to assign a custom key instead of a [TAB] or should change to something like [SHIFT + TAB] instead of TAB}''(GitHub \#7036). Eight users (7.2\%) expected Copilot to \textit{give suggestions when requested}. They did not want Copilot to suggest code all the time because it was interruptive, and they just wanted to get suggestions from Copilot when they requested. For example, one user asked ``\textit{Is it possible to not have GitHub Copilot automatically suggest code, instead only showing its suggestions when using the `trigger inline suggestion' shortcut?}'' (SO \#76147937). Seven developers expected Copilot to lanch \textit{a team version} and \textit{support access proxies}, respectively. \textit{Allow customization of the format of generated code} was mentioned by 5 developers. They wanted this feature to ``\textit{configure suggestion appearance}'' (GitHub \#7234) so that Copilot suggestions could be ``\textit{more distinguishable with normal code}'' and thus ``\textit{improve the accessibility}'' (GitHub \#7628). Few developers indicated that they only wanted to ``\textit{accept one line of several}'' of Copilot suggested code (SO \#75183662), and they called for the feature that they can \textit{accept the needed part of the suggestions}. Besides, four developers also mentioned that they hoped Copilot could be \textit{compatible with other code generation tools} like ReSharper.

\section{Implications}
\label{sec:implications}
This section presents the key implications of our study, which represent empirically grounded findings that could help researchers and developers to understand the effective usage of Copilot.

\noindent \textbf{Integration of Copilot with IDEs}: According to the results of RQ1.2 and RQ2.2, we found that most developers choose to integrate the Copilot plug-in with mainstream IDEs (including \textit{Visual Studio Code}, \textit{Visual Studio}, \textit{IntelliJ IDEA}, \textit{NeoVim}, and \textit{PyCharm}), and the percentage of mainstream IDEs used with Copilot by practitioners reaches 86.2\%. When developers choose the lesser known IDEs (e.g., \textit{Sublime Text}), they often find it hard to integrate the Copilot plug-in and thus have \textit{difficulty of integration}. In addition to the reason that developers may install Copilot in their chosen IDEs incorrectly, another reason for the \textit{difficulty of integration} is that Copilot does not support certain IDEs at the moment. When developers choose to use Copilot in mainstream IDEs, they can install it smoothly, and even if problems arise during the installation or use, they can easily find a solution on SO or GitHub Discussions as many other developers may have encountered similar issues. On the contrary, when developers choose to use Copilot in unpopular IDEs, they may not be able to install it because the IDEs are not officially supported by Copilot, and they cannot find solutions as few developers use Copilot with these IDEs. To reduce the \textit{difficulty of integration}, we recommend practitioners to use mainstream IDEs with Copilot. Besides, the most expected feature of users is that they hope Copilot \textit{can be integrated with more IDEs}, which confirms with the results of RQ2.2 (the main limitation of using Copilot is \textit{difficulty of integration}). We suggest GitHub can consider integrating Copilot with more IDEs in the future, which can meet the needs of diverse developers.

\noindent \textbf{Support for Front-end and Machine Learning Development}: As shown in the results of RQ1.1, RQ1.3, and RQ1.4, practitioners often write \textit{JavaScript} and \textit{Python} code when using Copilot, and they tend to use Copilot with front-end and machine learning related technologies (including frameworks, APIs, and libraries) to implement front-end (e.g., \textit{front-end element control}) and machine learning functions (e.g., \textit{data processing} and \textit{image processing}). \textit{JavaScript} is the foundation language of many popular front-end frameworks and most of Websites use \textit{JavaScript} on the client side. \textit{Python} is the first choice when it comes to the development of machine learning solutions with the help of rich libraries, e.g., \textit{OpenCV}. It is consequently reasonable that developers tend to use Copilot with \textit{JavaScript} to generate code for front-end and \textit{Python} for developing machine learning applications.

\noindent \noindent \textbf{Different Versions of Copilot}: From the results of RQ2.3, we can find that different versions of Copilot are needed (i.e., \textit{a team version}, \textit{a version for CLI (Command-Line Interface)}, and \textit{a on-premises version}). In different development environments, developers may have specific requirements when using Copilot. If GitHub can release different versions of Copilot, it would increase the usability and acceptance of Copilot and thus make it available to a wider variety of users. Besides, GitHub has officially launched some versions of Copilot, e.g., Copilot Labs, Copilot X, and Copilot Nightly. Copilot Labs~\cite{githubcopilotlabs} is used to experiment with new ideas before taking them into real production, Copilot X~\cite{githubcopilotx} provides an enhancement with new features, and Copilot Nightly contains experimental and less well tested changes. Developers can choose the version of Copilot according to their needs.

\noindent \textbf{Potentials and Perils of Using Copilot in Software Development}: Trained on billions of lines of code, Copilot can turn natural language prompts into coding suggestions across dozens of programming languages and make developers code faster and easier \cite{githubcopilot}. The results of RQ2.1 and RQ2.2 show that many benefits of using Copilot contradict its limitations and challenges, e.g., \textit{useful code generation} vs. \textit{limitation to code generation}. When deciding to use Copilot, developers should consider tool integration, user experience, budget, code privacy, and some other aspects, and make trade-offs between these factors. In short, using Copilot is like a double-edged sword, and practitioners need to consider various aspects carefully when deciding whether or not to use it. If Copilot can be used with appropriate programming languages and technologies to implement functions required by users correctly in developers' IDEs, it will certainly optimize developers' coding workflow and do what matters most - building software by letting AI do the redundant work. Otherwise, it will bring difficulties and restrictions to development, making developers feel frustrated and constrained. The study results can help practitioners being aware of the potential advantages and disadvantages of using Copilot and thus make an informed decision whether to use it for programming activities.

\noindent \textbf{Understanding the Code Generated by Copilot}: From the results of RQ2.1, RQ2.2, and RQ2.3, we can see that some practitioners think one of the benefits of using Copilot is that it has \textit{powerful code interpretation} feature, but some practitioners complained about \textit{the difficulty of understanding the generated code} by Copilot and called for \textit{code explanation} feature. It would be interesting to investigate why developers have opposing attitude towards understanding the code generated by Copilot and how the generated code by Copilot can be better explained to and understood by developers. Besides, GitHub Copilot Labs which is dependent on Copilot extension has been released for experimental purposes, and Copilot Labs has the feature to provide explanations of the code generated by Copilot for developers \cite{githubcopilotlabs}. The latest Copilot X also has the code explanation feature \cite{githubcopilotx}, but we do not know the reason why developers do not use Copilot Labs or Copilot X to interpret Copilot-generated code. We suggest that Copilot can provide the features for developers to better understand the generated code directly, such as generating code comments with the generated code in IDEs.

\noindent \textbf{Users' customization on suggestions by Copilot}: According to the results of RQ2.2, some developers thought that one of the limitations and challenges of Copilot is \textit{lack of customization}, and it is expected that a number of developers called for features of Copilot to \textit{allow customization of shortcuts for suggestions}, \textit{allow customization of the format of generated code}, and \textit{accept the needed part of the suggestions} (see the results of RQ2.3). The existing features of Copilot cannot satisfy users' needs on code suggestions. Users pointed out that they had difficulty in setting up shortcuts for actions on suggestions (e.g., accepting suggestions). Based on the feedback from users, they can only accept suggestions via the tab key, however, they want ``\textit{an option to change the keybinding for accepting the suggestions}'' instead (GitHub \#6919). Besides, a few users also reflected that it was hard to distinguish between the code wrote by themselves and the code suggested by Copilot. They wanted to customize the color, font, and format of Copilot suggestions. Another expected feature of users about Copilot is that they hope they can \textit{accept the needed part of the suggestions}. Some users want to accept Copilot suggestions line by line or accept only the next word of Copilot suggestions each time, while they do not want accept the entire suggestions by Copilot. As a result, there is a need for users to customize the way of accepting suggestions by Copilot. Some other expected features (e.g., \textit{allow setting filters for suggestions}, \textit{suggestions in IDE UI can be configured}, and \textit{ability to select the training sources for suggestions}) also relate to customization on suggestions by Copilot. On the basis of the above feedback from users, we believe that it is necessary for Copilot to allow customization for suggestions, which will give developers better user experience when using Copilot as they are able to use it in the way they want.

\noindent \textbf{Towards an Effective Use of Copilot}: Further investigation about the practices of Copilot can be conducted by questionnaire and interview. Under what conditions the challenges of using Copilot will show up as advantages or disadvantages, and how to use Copilot to convert its disadvantages into advantages are also worth further exploration. Besides, although we have investigated various aspects of using Copilot (e.g., limitations and challenges), we did not looked in depth at what types of users (e.g., developers, educators, and students) who use Copilot, when and how they use Copilot. By exploring these aspects, researchers can get meaningful information which would help guide towards an effective use of Copilot.

\section{Threats to Validity}
\label{sec:threats}
The validity threats are discussed according to the guidelines in \cite{runeson2009guidelines}, and internal validity is not considered, since we did not investigate the relationships between variables and results. The three validity threats presented below highlight potential limitations of this research that may invalidate some of the results. Future research can focus on minimizing these threats to ensure methodological rigor of the study.

\textbf{Construct validity} indicates whether the theoretical and conceptual constructs are correctly measured and interpreted. We conducted data labelling, extraction, and analysis manually, which may lead to personal bias. To reduce this threat, the data labelling of SO posts was performed after the pilot labelling to reach an agreement between the authors. The data extraction and analysis was also conducted by two authors, and the first author rechecked all the results produced by the third author. During the whole process, the first author continuously consulted with the second author to ensure there are no divergences.

\textbf{External validity} indicates the the degree of generalization of the study results, i.e., the extent to which the results can be generalized to other contexts. We chose two popular developer communities (SO and GitHub Discussions) because SO has been widely used in software engineering studies and GitHub Discussions is a new feature of GitHub for discussing specific topics \cite{hata2022github}. These two data sources can partially alleviate the threat to external validity. However, we admit that our selected data sources may not be representative enough to understand all the practices, challenges, and expected features of using Copilot.

\textbf{Reliability} indicates the replicability of a study yielding the same or similar results. We conducted a pilot labelling before the formal labelling of SO posts with two authors, and the Cohen's Kappa coefficient is 0.773, indicating a decent consistency. We acknowledge that this threat might still exist due to the small number of posts used in the pilot. All the steps in our study, including manual labelling, extraction, and analysis of data were conducted by three authors. During the process, the three authors discussed the results until there was no any disagreements in order to produce consistent results. In addition, the dataset of this study that contains all the extracted data and labelling results from the SO posts and GitHub discussions has been provided online for validation and replication purposes \cite{replpack}.

\section{Conclusions}
\label{sec:conclusions}
We conducted an empirical study on SO and GitHub discussions to understand the practices, challenges, and expected features of using GitHub Copilot from the practitioners' perspective. We used ``\textit{copilot}'' as the search term to collect data from SO and collected all the discussions under the ``Copilot'' category in GitHub discussions. Finally, we got 303 SO posts and 927 GitHub discussions related to Copilot. Our results identified the programming languages, IDEs, technologies used with Copilot, functions implemented by Copilot, and the benefits, limitations, and challenges of using Copilot, which are first-hand information for developers. The main results are that: (1) \textit{JavaScript} and \textit{Python} are the most frequently discussed programming languages by developers with Copilot. (2) \textit{Visual Studio Code} is the dominant IDE used with Copilot. (3) \textit{Node.js} is the major technology used with Copilot. (4) \textit{Data processing} is the main function implemented by Copilot. (5) \textit{To help generate code} is the leading purpose of users using Copilot. (6) \textit{Useful code integration} is the most common benefit mentioned by developers when using Copilot. (7) \textit{Difficulty of integration} is the most frequently encountered limitations and challenges when developers use Copilot. (8) Copilot \textit{can be integrated with more IDEs} is the most expected feature of users.

In the next step, we plan to explore the practices of using Copilot by conducting interviews or an online survey to get practitioners' perspectives on using Copilot, which can supplement our existing data collected from repository mining. Besides, we also plan to further explore various aspects of Copilot, especially how to improve the understanding of developers on the generated code (see Section \ref{sec:implications}).

\section*{Acknowledgements}
This work has been supported by the Natural Science Foundation of China (NSFC) under Grant No. 62172311 and the Special Fund of Hubei Luojia Laboratory.

\bibliographystyle{IEEEtran}
\bibliography{References}
\end{document}